\documentclass[aps,prd,showpacs]{revtex4}
\usepackage{graphicx}
\usepackage{epsfig}
\usepackage{psfig}
\usepackage{revsymb}
\begin{document}
\renewcommand{\thesection}{\arabic{section}}
\renewcommand{\thesubsection}{\arabic{subsection}}
\title{The Study of Magnetically Deformed Atoms in the Outer Crust of Neutron Stars in Presence of Strong
Quantizing Magnetic Field}
\author{Arpita Ghosh and Somenath Chakrabarty$^\dagger$}
\affiliation{
Department of Physics, Visva-Bharati, Santiniketan 731 235, 
West Bengal, India\\ 
$^\ddagger$E-mail:somenath.chakrabarty@visva-bharati.ac.in}
\begin{abstract}
We have studied the various properties of magnetically deformed atoms, replaced by deformed Wigner-Seitz
cells, at the outer crust region of strongly magnetized neutron stars (magnetars) using a relativistic 
version of Thomas-Fermi model in cylindrical coordinates. 
\end{abstract}
\maketitle
\section{Introduction}
From the observational evidence of a few strongly magnetized neutron stars, which are the sources of
anomalous X-rays and soft gamma rays, also called magnetars \cite{R1,R2,R3,R4}, the study of crustal matter, 
in particular the outer crust of such compact stellar objects have gotten a new dimension. These exotic 
objects are also called anomalous X-ray pulsars (AXP) and soft gamma repeaters (SGR). The outer crust of a 
typical neutron star in general, is mainly composed of dense crystalline metallic iron. The density of such 
metallic crystalline matter is $\sim 10^{-4}\rho_0$, where $\rho_0$ is the normal nuclear density 
($\sim 2.8\times 10^{14}$gm/cc). Therefore, it is absolutely impossible 
to investigate the properties of such ultra-dense matter in material
science laboratories, even at zero magnetic field. The observed surface magnetic field of the magnetars
is $\sim 10^{15}$G, which is again too high to achieve in the terrestrial laboratories. Also, it is quite 
possible that the interior field of such exotic objects can go up to $\sim 10^{18}$G (which can be
shown theoretically by Virial theorem). If the magnetic field at the interior is really so high, then most 
of the physical and chemical properties of
the dense neutron matter should change significantly from the conventional neutron star (radio
pulsars) scenario (see the recent article by one of the co-authors of this article \cite{R5} for necessary 
references). Even though the magnetic field strength at the crustal region is slightly higher than  
$10^{15}$G, it must change significantly most of the properties of dense matter, both in the outer crust and 
the inner crust regions of the magnetars \cite{R5,R6}. It is believed that
strong magnetic field can cause a structural deformation of the metallic atoms present in the outer
crust of a neutron star. The spherical symmetry of the atoms will be destroyed and becomes cigar shape with 
the elongated axis along the direction of strong magnetic field. The atoms may even become almost an one 
dimensional string like object, i.e., needle shape, if the magnetic field strength is extremely strong. In a 
future article we shall present the problem related to structural deformation of atoms in a strong quantizing
magnetic field using a completely different approach \cite{RR6}.
                                           
In this article we shall study the properties of outer
crust matter composed of magnetically deformed metallic (iron) atoms. In section 2 we have developed
the basic formalism and discuss the numerical results, whereas in the last section we
have given the conclusion and the future prospect of this work. In this article, for the sake of
simplicity, we shall assume a cylindrical deformation of the atoms in the outer crust region and use
cylindrical coordinate system with azimuthal symmetry. In reality, to investigate the cigar like
deformed atoms in presence of strong magnetic field one has to use prolate spheroidal coordinate
system.  
\section{Basic formalism}
The width of the outer crust of a typical neutron star is $\sim 0.3$ km, the density of matter, which is
assumed to be a dense crystal of metallic iron is $\sim 4.3\times 10^{11}$gm/cc \cite{SHA}. To investigate the
properties of such dense exotic crystalline matter of metallic iron, we have
replaced the outer crust matter by a regular array of cylindrically deformed (in our future work we
shall consider more realistic geometrical structure of the deformed atoms, which is prolate spheroidal
in nature) Wigner-Seitz (WS) cells, 
with the positively charged nucleus at the centre surrounded by a non-uniform electron gas. The axis of
each cylinder is along the direction of magnetic field and further assumed azimuthal symmetry for all the
cylindrically deformed WS cells.

We start with the Poisson's equation, given by
\begin{equation}
\nabla^2\phi=4\pi en_e
\end{equation}
where $\phi$ is the electrostatic field, e is the electron charge and $n_e$ is the electron density,
which because of assumed non-uniformity within the WS cell,  is a function of positional coordinates $(r,z)$.
Now in the cylindrical coordinate with circular symmetry, the above equation reduces to
\begin{equation}
\frac{\partial^2\phi}{\partial r^2}+\frac{1}{r}\frac{\partial\phi}{\partial
r}+\frac{\partial^2\phi}{\partial z^2}=4\pi en_e
\end{equation}
It is well known that in the presence of strong quantizing magnetic field, the number density of
degenerate electron gas is given by 
\begin{equation}
n_e=\frac{eB}{2\pi^2}\sum_{\nu=0}^{\nu_{max}}(2-\delta_{\nu 0})p_F
\end{equation}
where B is the constant external magnetic field, assumed to be acting along Z-direction and is $>B_c
^{(e)}$, where $B_c
^{(e)}$ is the typical strength of magnetic field beyond which, in the relativistic region the Landau
levels for the electrons are populated. For the sake of convenience, throughout this article we shall
use $\hbar=c=1$. The critical strength is given by $B_c=m_e ^2/\mid e\mid$ \cite{R5} and assume that the 
matter is at zero temperature,
where $m_e$ is the electron rest mass and $\mid e\mid$ is the magnitude of electron charge. In eqn.(3), 
$p_F$ is the electron Fermi momentum, $\nu$ is the Landau quantum
number, with $\nu_{max}$, the upper limit of $\nu$. The factor $(2-\delta_{\nu 0})$ takes care of
singly degenerate $\nu=0$ state and doubly degenerate all other states with $\nu\neq 0$.
To study the properties of outer crust matter with deformed WS cells, we make Thomas-Fermi
approximation, which is a semi-classical approach and having a lot of shortcomings and limitations
\cite{R7}. We further use the Thomas-Fermi condition, given by
\begin{equation}
\mu_e= (p_F ^2+m_e ^2+2\nu eB)^{1/2}-e\phi={\rm{constant}}
\end{equation}
where $\mu_e$ is the electron chemical potential and in this model, it is assumed to be constant
throughout the WS cell.
From this equation we can express the Fermi momentum of electrons in the following form: 
\begin{equation}
p_F=[(\mu_e+e\phi)^2-m_e ^2-2\nu eB]^{1/2}
\end{equation}
Since the electrostatic potential $\phi\equiv \phi(r,z)$, the Fermi momentum $p_F$ for electron is also a 
function of positional coordinates $(r,z)$ within the cell. In principle
one should use the exact expression for electron Fermi momentum as given above, in the equation for 
electron density (eqn.(3)) which is in turn appearing on the right hand side of the
cylindrical form of Poisson's equation (eqn.(2)). However, with this exact expression for $p_F$, it is
absolutely impossible to proceed further analytically, even a single effective step. From the very
beginning, therefore, one has to use some numerical technique to solve the Poisson's equation, which
is an extremely complicated nonlinear partial differential equation. Of course, with the numerical method 
within the limitation of the algorithm followed, we will get more exact results.
However, in numerical computation of $\phi(r,z)$, we get a set of numbers, but the beauty of this model
will be completely destroyed and a lot of interesting physics associated with intermediate results of this 
problem will be totally lost. In this context, let us quote from a
quite old but extremely interesting paper by Wigner and Bardeen \cite{R8}- "It is perhaps not quite 
superfluous to have, in addition to a more exact  calculations of a physical quantity, an approximate 
treatment which merely shows how the quantity in question is determined, and the lines along which a more 
exact calculation could be carried out. Such a treatment often leads to a simple formula by means of
which the magnitude of the quantity may be readily determined". Therefore, to get an approximate analytical
solution for $\phi(r,z)$, we assume 
that $\nu_{max}=0$ and neglect the rest mass of electron, i.e., we put $m_e=0$
in the expression for Fermi momentum $p_F$ (eqn.(5)).  The approximation $\nu_{max}=0$ is actually valid if
the magnetic field is extremely high. However, to investigate the properties of dense electron gas
within the cylindrically deformed cells in a little bit of exact manner later in this article, we
shall use the approximate solution for $\phi(r,z)$, but do not restrict ourselves to $\nu_{max}=0$. 
In fact we shall show later that the upper limit 
$\nu_{max}$ for the electron Landau quantum number is also a function of $(r,z)$. Therefore to evaluate 
various
physical quantities in the outer crust region, we have to first obtain  an approximate solution for the
Poisson's equation. To achieve our goal, we use the approximate form of electron Fermi momentum
obtained from the assumption as mentioned above and its mathematical form is  given by
\begin{equation}
p_F \sim \mu_e+e\phi
\end{equation}
Next on substituting
\begin{equation}
\mu_e+e\phi(r,z)=\psi(r,z),
\end{equation}
the cylindrical form of Poisson's equation reduces to
\begin{equation}
\frac{\partial^2\psi}{\partial r^2}+\frac{1}{r}\frac{\partial\psi}{\partial
r}+\frac{\partial^2\psi}{\partial z^2}=\lambda^2\psi
\end {equation}
where
$\lambda^2=2e^3 B/\pi$. Under this approximation, the Poisson's equation, as shown above, reduces to a
linear partial differential equation.  To solve this partial differential equation analytically we 
use the method of separation of variables, given by 
\begin{equation}
\psi(r,z)=R(r)Z(z)
\end{equation} 
Substituting $\psi(r,z)$ from eqn.(9) in eqn.(8) and introducing a constant $\xi$, we get
\begin{eqnarray}
&&\frac{d^2 R}{dr^2}+\frac{1}{r}\frac{dR}{dr}+\xi^2 R=0\\
&&\frac{d^2 Z}{dz^2}-(\xi^2+\lambda^2)Z=0
\end{eqnarray}
where $\xi$ is some real constant, independent of r and z but may change with the magnetic field
strength and with the mass number and the atomic number of the type of elements present in the outer
crust region. The solutions of eqns.(10) and (11) are well known. For eqn.(10), the solution is a Bessel
function of order zero with the argument $\xi r$, whereas for eqn.(11), it is an exponentially
decaying function of $z$. In the language of mathematics,    
the solution for $\psi(r,z)$ is then given by
\begin{equation}
\psi(r,z)=CJ_0(\xi r)exp\left [\pm(\xi^2+\lambda^2)^{1/2}z\right ]
\end{equation}
where $+$ and $-$ signs are for $z<0$ or $>0$ respectively. Here $C$ is a constant (again may change with 
the magnetic field strength and with
the atomic properties of the elements present in the outer crust) and $J_0(\xi r)$
is the Bessel function of order zero.
Now on the nuclear surface, at the centre of the WS cells, $\phi=Z_0 e/r_n $ \cite{R9}, where $Z_0$ is the 
atomic number and $r_n=r_0 A^{1/3} $ is the nuclear radius, $r_0=1.12$fm and $A$ is the mass number (here we 
have assumed that the nuclei remain spherical, even if the magnetic field is too high). For
the sake of simplicity we put $z=0$ and $r=r_n$ on the nuclear surface, i.e., we have chosen points on
nuclear surface along $r$-axis, which gives
\begin{equation}
\psi(r_n,0)=CJ_0(\xi r_n)=\mu_e+\frac{Z_0e^2}{r_n}
\end{equation}
Hence we get
\begin{equation}
C=\frac{1}{J_0(\xi r_n)}\left [\mu_e+\frac{Z_0e^2}{r_n}\right ]
\end{equation}
In the expression, of course, $\xi$ is still an unknown quantity. To determine $\xi$, we now put
$z=+r_n$ and $r=0$ in the expression for $\phi(r,z)$ on the nuclear surface, i.e., we have chosen a point on 
the positive side of $Z$-axis. 
Then using $J_0(0)=1$, we have (because of the symmetry about $z=0$ plane, $z=-r_n$ and $r=0$ will also give 
the same result)
\begin{equation}
\psi(0,r_n)=C \exp\left [-(\xi^2+\lambda^2)^{1/2}r_n\right ]=\mu_e+\frac{Z_0 e^2}{r_n}
\end{equation}
Hence we get 
\begin{equation}
\exp\left [-(\xi^2 +\lambda^2)^{1/2}r_n\right ]=J_0(\xi r_n)
\end{equation}
This is a  highly transcendental equation for $\xi$. However, it is possible to evaluate $\xi$ numerically 
from this equation for a given magnetic field strength and for a given type of element, say metallic 
iron in the crystalline form present in the outer crust region. The above
expression puts some constraint on the range of $\xi r_n$. To make this point more transparent let us
express eqn.(16) in the following form
\begin{equation}
\xi^2+\lambda^2=\frac{1}{r_n^2}[\ln J_0(\xi r_n)]^2
\end{equation}
Since $J_0(\xi r_n)$ is an oscillatory function and becomes negative as $\xi r_n$ crosses the first
zero, in the numerical computation we therefore put a check in the computer code to restrict $\xi r_n$
from $0+$ to the first zero$-$, which is $2.40482......$.
In fig.(1) we have plotted $\xi$ (in MeV) as a function of magnetic field strength $B$, expressed in terms of
critical magnetic field strength $B_c^{(e)}$. The variation is insensitive for the low
and moderate values of magnetic field strengths. However, the overall magnitude of $\xi$ is very close
to $100$MeV. This figure shows that beyond field strength $10^{17}$G, $\xi$ increases sharply. In
fig.(2) we have plotted the variation of normalization constant $C$ with the magnetic field
strength, expressed in the same unit as in fig.(1). For low and moderate magnetic 
field values it is almost constant, then it falls abruptly beyond
$10^{17}$G and finally saturates to a constant value equal to $\sim 100$MeV. Since beyond $10^{17}$G,
electrons within the cells occupy their zeroth or very low lying Landau levels, the quantum mechanical
effect of magnetic field becomes extremely important and as a consequence both $\xi$ and $C$ change
significantly beyond this magnetic field value.

Since the WS cells are overall charge neutral, then at any point $(r,z)$ on the surface, even on the plane
faces or on the boundary points of $z=0$ plane, we have in cylindrical coordinate 
\begin{equation}
\nabla\psi=0
\end{equation}
This condition gives 
\begin{equation}
\hat{e_r}\frac{\partial\psi}{\partial r}+ \hat{e_z}\frac{\partial\psi}{\partial z}=0
\end{equation}
which must be satisfied at all the surface points on the WS cylinder.

Now from the Thomas-Fermi condition, we have
\begin{equation}
p_F=[\psi^2(r,z)-m_{\nu}^2]^{1/2}
\end{equation}
where $m_{\nu}=(m_e^2+2\nu eB)^{1/2}$.
With this more exact expression for $p_F$, the number density for electron gas can be expressed as 
\begin{equation}
n_e(r,z)=\frac{eB}{2\pi^2}\sum_{\nu=0}^{\nu_{max}}(2-\delta_{\nu 0})[\psi^2(r,z)-m_{\nu}^2]^{1/2}
\end{equation}
This is obviously more exact than eqn.(3). Further, this expression shows that the electron density
is a function of both $r$ and $z$ within the WS cell. Which justifies the assumption that the electron
distribution inside each WS cell around the fixed nucleus is non-uniform.
Now from the non-negative nature of $p_F ^2$, we have
\begin{equation}
\nu_{max}=\frac{\psi^2(r,z)-m_e ^2}{2eB}=\nu_{max}(r,z)
\end{equation}
The upper limit of Landau quantum number $\nu_{max}$ will therefore also depend on the positional 
coordinates of the associated electron within the cell.
In fig.(3) we have plotted electron number density in terms of normal nuclear density multiplied by
$10^4$, as a function of
radial distance from nuclear surface to the WS cell boundary for the magnetic field strengths $10^1$,
$5\times 10^2$, $5\times 10^3$, $10^4$ and $5\times 10^4$ times $B_c^{(e)}$, indicated by the curves 
$a$, $b$, $c$, $d$ and
$e$ respectively. These curves show that the electron number density is maximum near the nuclear
surface and minimum near the WS cell boundary $r_{max}$. This figure also shows that the value of electron
number density increases with the strength of magnetic field. In fig.(4) we have plotted the same
quantity as in fig.(3) but against the axial distance $z$ from the nuclear surface to the WS boundary
indicated by $z_{max}$.
In this case also the variations are exactly same as in fig.(3). The qualitative difference is because
of different types of functional dependence.
In fig.(5) we have plotted the upper limit of Landau quantum
number $\nu_{max}$ as a function of radial coordinate from the nuclear surface to the WS cell 
boundary. In this figure
the upper curve is for $10\times B_c^{(e)}$, the middle one is for $10^2\times B_c^{(e)}$ and the lower 
one is for 
$500\times B_c^{(e)}$. The value of $\nu_{max}$ decreases with the increase in magnetic field strengths.
It has been observed that beyond $500 \times B_c^{(e)}$, $\nu_{max}$ becomes identically zero throughout 
the WS cell. Further, the value of $\nu_{max}$ for $B\leq 500 \times B_c^{(e)}$ is largest near the nuclear
surface and exactly zero at the cell boundary. In fig.(6) we have plotted the same kind of variation
of $\nu_{max}$, but against the axial distance from nuclear surface to the cell boundary. The
qualitative and the quantitative variations are exactly identical with fig.(5). These two figures show that
the electrons are completely spin polarized in the direction opposite to the magnetic field $\vec B$
at the cell boundary for $B\leq 500 \times B_c^{(e)}$, but beyond this value they are polarized at all
the points within the cell.
Although the variations along radial and axial directions are shown in these two figures, we expect
that such polarized picture of electron gas will be there throughout the cylindrically deformed
WS cell surface, including the two plane faces.

Now the minimum possible value of $\nu_{max}$ is zero, hence we have
\[
\psi^2 (r,z)-m_e ^2 \geq 0
\]
To get the length and radius of the cylindrically deformed WS cell, we first put $r=0$ in the above
expression. Then we have 
\begin{equation}
C^2 \exp[\pm 2z(\lambda^2+\xi^2)]\geq m_e ^2
\end{equation}
Therefore beyond some value of $z$, say $z_{max}$, this inequality will break, i.e., just at 
value $z=z_{max}$, we have
\begin{equation}
C^2 \exp[\pm2z_{max}(\lambda^2+\xi^2)^{1/2}]=m_e ^2
\end{equation}
Since $z$-axis is symmetric about $z=0$ plane, we have from this relation
\begin{equation}
z_{max}=\frac{1}{(\lambda^2+\xi^2)^{1/2}}\mid\ln\left ( \frac{m_e}{C}\right ) \mid
\end{equation}
This is the magnitude of $z$ on the plane faces of a WS cell, i.e., $2z_{max}$ is the length of the
cylindrically deformed WS cell for a given B and a set of $(Z_0,A)$. In fig.(7) we have shown the
variation of $z_{max}$ with the strength of magnetic field $B$ expressed as before in terms of critical field
strength $B_c^{(e)}$. This figure shows that the variation is
almost insensitive for low and moderate magnetic field strengths but decreases almost abruptly beyond
$10^{16}$G, when most of the electrons occupy their zeroth Landau level, and finally 
tends to saturate to a constant value $\sim 10$fm.

To get the radius of such cells, we next put $z=0$, then we have
\[
\psi^2(r,0)-m_e ^2 \geq 0
\]
which further gives for $r=r_{max}$, the radius of the cylindrical WS cells, 
\begin{equation}
J_0(\xi r_{max})=\frac{m_e}{C}
\end{equation}
Obviously an analytical solution for $r_{max}$ is absolutely impossible, we obtain numerically the value
of $r_{max}$ for a given $B$ and $(A,Z_0)$. In this case also, as before, we
restrict the value of the product $\xi r_{max} <$ the first zero of $J_0(\xi r_{max})$. The variation
of $r_{max}$ with the strength of magnetic field is shown in fig.(8). The nature of variation is more
or less same as
that of $z_{max}$. However, we have noticed that for extremely large field strength, 
$r_{max}\longrightarrow 0$.
This is a remarkable difference from its longitudinal counter part. It actually shows that in presence
of extremely strong magnetic field the cylinders become more and more thin in the transverse
direction. We therefore conclude that with the increase in magnetic field
strength the radial contraction will be enormous compared to the axial one. From figs.(7)-(8) we have noticed
that the variations are most significant  beyond $B=10^{16}$G. The reason is again because of the fact that
the electrons occupy only their zeroth Landau level in presence of such strong magnetic field, 
at which the quantum mechanical effect of the magnetic field dominates.

Next we calculate the different kinds of energies of electron gas within the WS cells. The cell
averaged kinetic energy density of an electron is given by
\begin{eqnarray}
\epsilon_k=\frac{2}{V}\frac{eB}{2\pi^2}\int d^3r&&\sum_{\nu=0}^{\nu_{max}} (2-\delta_{\nu 0}) \nonumber \\ &&
\int_0^{p_F(r,z)} dp_z [(p_z^2+m_\nu^2)^{1/2}-m_e]
\end{eqnarray}
where in the cylindrical coordinate system with azimuthal symmetry $d^3r=2\pi rdr dz$, with the limits
$r_n \leq r \leq r_{max}$ and $r_n \leq z \leq z_{max}$ and $V=\pi r_{max}^2 2\times z_{max}$, the
volume of each cell. The
$p_z$ integral is trivial and can be obtained an analytic expression for the average kinetic energy
density. However, if we do not integrate
over $r$ and $z$, then from this equation we get the energy density $\epsilon_k(r,z)$ at a 
particular point inside the cell.  Of course, in that
case we need not have to divide by $V$. The factor $2$ is for $z$-symmetry about $z=0$ plane. In
fig.(9) we have plotted the variation of kinetic energy density as a function of radial distance $r$ from the
nuclear surface to the cell boundary, keeping $z=0$. This figure shows that the kinetic energy density
increases with the
increase in magnetic field strength. We have indicated the curves by $a$, $b$, $c$, $d$ and $e$ for
the magnetic field strengths $10B_c^{(e)}$, $5\times 10^2B_c^{(e)}$, $5\times 10^3B_c^{(e)}$, 
$10^4B_c^{(e)}$ and
$5\times 10^4B_c^{(e)}$ respectively. In fig.(10) we have shown the same kind of variation along axial
distance from the nuclear surface to one of the plane faces of the WS cell. The variation with
magnetic field strength is again almost identical with fig.(9). Similar to the variation of electron number  
density within the cell (fig.(3) and fig.(4)), both fig.(9) and fig.(10) show that the
kinetic energy density for electrons is maximum near the nuclear surface and minimum at the cell boundary. 

Similarly, the cell averaged electron-nucleus interaction
energy per unit volume is given by 
\begin{equation}
E_{en}=-\frac{2}{V} Z_0e^2 \int d^3r \frac{n_e(r,z)}{(r^2+z^2)^{1/2} }
\end{equation}
Analogous to the kinetic energy, here also one can obtain the interaction energy per unit
volume, $E_{en}(r,z)$ at a particular point within the cell. In fig.(11), the variation of the magnitude of
electron-nucleus interaction energy per unit volume with the radial distance is plotted. In this figure 
we  have indicated the curves by $a$, $b$, $c$, $d$ and $e$ for
the magnetic field strengths $10B_c^{(e)}$, $5\times 10^2B_c^{(e)}$, $5\times 10^3B_c^{(e)}$, 
$10^4B_c^{(e)}$ and
$5\times 10^4B_c^{(e)}$ respectively. This figure shows that the magnitude of electron-nucleus 
interaction energy increases with the increase in magnetic field strength. Which means that
with the increase in magnetic field strength the electrons become more strongly bound by
the nuclear Coulomb attractive potential. 
It is also obvious from the figs.(9)-(12) that for a given magnetic field
strength the kinetic energy density and the magnitude of electron-nucleus interaction energy per unit
volume at a particular point, either along axial direction or in the radial direction,  inside the WS
cell are of the same order of magnitude. We do believe that this is true at all the points inside the
WS cell. In fig.(12) we have shown the same kind of variation along z-axis. Both the qualitative and the
quantitative nature of variations with the strength of magnetic field are same as that of fig.(11). 

Next we consider the cell averaged electron-electron direct interaction energy density, given by
\begin{equation}
E_{ee}^{dir}=\frac{1}{V}e^2\int d^3r n_e(r,z)\int d^3r^\prime n_e(r^\prime,z^\prime) \frac{1}{[(\vec r
-\vec r^\prime)^2 +(z-z^\prime)^2]^{1/2}}
\end{equation}
Now assuming $\vec r$ as the reference axis, we have $(\vec r-\vec r^\prime)^2 =r^2+{r^\prime}^2
-2rr^\prime\cos \theta$, where $\theta$ is the angle between $\vec r$ and $\vec r^\prime$. Then
$d^3r^\prime=r^\prime dr^\prime d\theta dz^\prime$, with $0\leq \theta \leq 2\pi$. 
In this case the $z$ integral has to be broken into two parts, one with limit $-z_{max} \leq z \leq
\overline z$ and the other one with the limit $\overline z \leq z \leq +z_{max}$. The value of
$\overline z$ is not easy to evaluate in the region between $Z$-axis and $r$-axis. With $\theta$
symmetry, for the sake of simplicity we put $\overline z=r_n$ and expect that
the error will be nominal. Now to obtain electron-electron
direct interaction energy one has to evaluate the five dimensional integral as shown in eqn.(29). 
None of them can be obtained
analytically, hence it is necessary to follow some numerical method. Even the $\theta$ integral can
not be obtained analytically. One can express the $\theta$ integral in the form of an elliptical
integral of first kind, given by
\begin{equation}
I_{\cal{EL}}(r,r^\prime,z,z^\prime)=\int_0^{\pi/2} d\theta \frac{1}{(1-K\cos^2\theta)^{1/2}}
\end{equation}
where $K=4rr^\prime/[(r+r^\prime)^2+(z-z^\prime)^2]$. The direct part is then given by
\begin{equation}
E_{ee}^{dir}=\frac{4}{2V}e^2\int d^3r n_e(r,z)\int r^\prime dr^\prime dz n_e(r^\prime,z^\prime) \frac{1}{[(r
+ r^\prime)^2 +(z-z^\prime)^2]^{1/2}} I_{\cal{EL}}(r,r^\prime,z,z^\prime)
\end{equation}
where the factor $4$ is coming from the angular integral over $\theta$ from $0$ to $2\pi$.
Let us now consider the
elliptic integral (eqn.(30)) on $z=0$-plane. In this case both $z$ and $z^\prime$ are zero and the
factor $K=1$. Then it can be shown very easily that the integral given by eqn.(30) will diverge at the lower
limit. To avoid this unphysical infinity we put a lower cut off (infrared cut off) $\delta$, which
will now the lower
limit for $\theta$. The physical meaning of non-zero lower limit for the $\theta$-integral is that
the two electrons under consideration can not be at zero distance from
each other on the arc of a circle whose centre is same as that of the nucleus. The infrared cut off
$\delta$ which is a measure of angular distance between two neighboring electrons 
must necessarily depends on the minimum possible linear distance between them and also on the
radial distance from the centre. From a very elementary geometrical construction it can be shown that 
\begin{equation}
\delta=\frac{s}{r}
\end{equation}
where $s$ is the arc length, or the distance between two neighboring electrons on the circular arc. 
Since $s$ is infinitesimal in nature, we can approximate it by a straight
line of length $s$, which is the length of the cord connecting two points occupied by two neighboring
electrons. Since $s$ is the minimum possible distance between two electrons, we can express
it in terms of electron density near those points, given by 
\begin{equation}
s\sim n_e^{1/3}
\end{equation}
In a future publication we shall evaluate this multidimensional integrals along with the
electron-electron exchange energy (as given below) using the spinor solutions of Dirac equation in cylindrical
coordinate in presence of strong quantizing magnetic field using a sophisticated Monte-Carlo numerical code
for multidimensional integrals \cite{ARP}. 

The electron-electron exchange energy corresponding to the $i$th electron in the cell is given by
\begin{equation}
E_{ee}^{(ex)}= -\frac{e^2}{2}\sum_j \int d^3r d^3r^\prime \frac{1}{ [(\vec
r-\vec {r^\prime})^2+(z-z^\prime)^2]^{1/2}} \bar\psi_i(\vec r,z)\bar\psi_j(\vec {r^\prime},z^\prime)
\psi_j(\vec r,z)\psi_i(\vec {r^\prime},z^\prime)
\end{equation}
where $\psi_i(r,z)$ is the spinor wave function in cylindrical coordinate in presence of strong
quantizing magnetic field, and $\bar \psi({\vec r},z)=
\psi^\dagger({\vec r},z)\gamma_0$, the adjoint of the spinor and $\gamma_0$ is
the zeroth part of the Dirac gamma matrices $\gamma_\mu$ in cylindrical coordinate system.

The kinetic pressure of non-uniform electron gas within the WS cell is given by 
\begin{equation}
P(r,z)=\frac{eB}{2\pi^2}\sum_{\nu=0}^{\nu_{max}(r,z)}(2-\delta_{\nu 0})\int_0^{p_F(r,z)}
\frac{p_z^2dp_z}{(p_z^2+m_\nu^2)^{1/2}} 
\end{equation}
The $p_z$ integral is very easy to evaluate analytically and is given by 
\begin{eqnarray}
P(r,z)&=&\frac{eB}{4\pi^2}\sum_{\nu=0}^{\nu_{max}(r.z)}\big [p_F(p_F^2+m_\nu^2)^{1/2}\nonumber \\ &-& m_\nu^2
\ln\left \{ \frac{p_F+(p_F^2+m\nu^2)^{1/2}}{m_\nu}\right \}\big ]
\end{eqnarray}
This equation shows that the electron kinetic pressure also changes from point to point within the WS
cells. In fig.(13) we have shown the variation of kinetic pressure for the non-uniform electron gas with the
radial distance within the cell. Curves $a$ and $b$ are for $B=10\times B_c^{(e)}$ and 
$B=500\times B_c^{(e)}$, whereas upper curve and lower curve as indicated by $c$ (almost identical) are for   
$B=5000\times B_c^{(e)}$ and $B=10000\times B_c^{(e)}$ respectively. In fig.(14) the same kind of
variations are shown along z-axis. In this figure the curves for 
$B=5000\times B_c^{(e)}$ and $B=10000\times B_c^{(e)}$ are almost identical and indicated by single thick
curve $c$. For $B=10\times B_c^{(e)}$ and 
$B=500\times B_c^{(e)}$ the curves are indicated by $a$ and $b$ respectively. These two figures show
that the kinetic pressure is maximum near the nuclear surface and zero at the cell boundary. The
variation with magnetic field strength shows that the non-uniform electron gas becomes softer for high
magnetic field.
\section{Conclusions}
In this article we have investigated various physical properties of non-uniform electron gas within
the cylindrically deformed
atoms of metallic iron at the outer crust of a strongly magnetized neutron star (magnetar).
Because of extremely strong magnetic field, we
have assumed a cylindrical type deformation of the atoms, replaced by WS cells with the same kind of
deformation. The axes of the
cylinders are along the direction of magnetic lines of forces. The curved surfaces of these cylinders
are therefore parallel to the boundary surface of the neutron stars in the region far away from the
magnetic poles. We have studied the variation of electron
number density, kinetic energy density, electron-nucleus interaction energy per unit volume of
electron gas and also kinetic pressure of non-uniform electron gas within the cylindrically deformed
WS cells with the strength of magnetic field  for a given spatial coordinate ($r,z$) and 
also with the radial and axial distances within the cell for a given magnetic field value. We have also
investigated the variations of longitudinal and transverse dimensions of deformed WS cells with the
strength of magnetic field. We have noticed that the transverse dimension of a cylindrically deformed
WS cell becomes extremely thin in presence of ultra-strong magnetic field. It is believed that the 
atoms become cigar shape in presence of strong magnetic field. In our future study, the properties of 
neutron star crustal matter with cigar
shape atoms in the metallic crystal in presence of strong magnetic field. In future, we shall also
evaluate the electron-electron direct interaction energy and the exchange part of electron-electron
interaction.

\begin{figure}[ht]
\psfig{figure=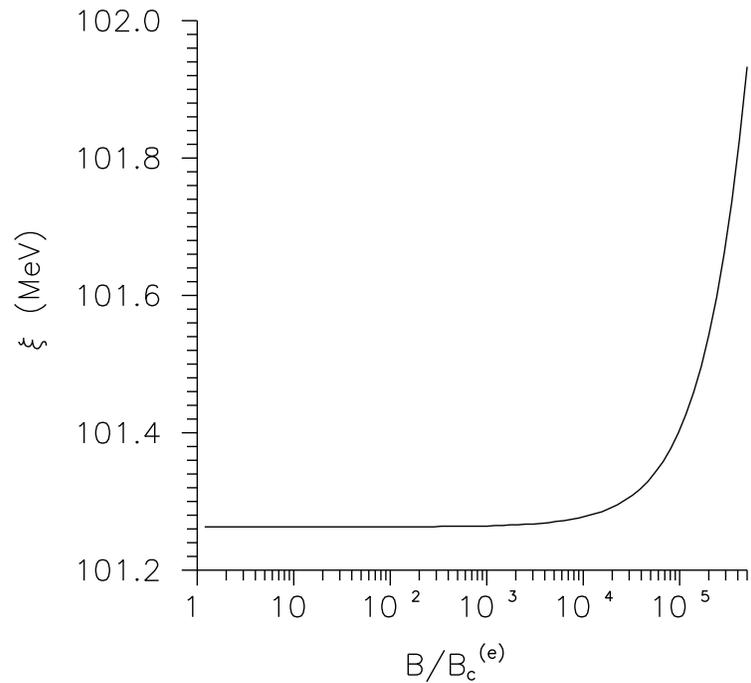,height=0.5\linewidth}
\caption{The variation of separation variable $\xi$ with the magnetic field strength }
\end{figure}
\begin{figure}[ht]
\psfig{figure=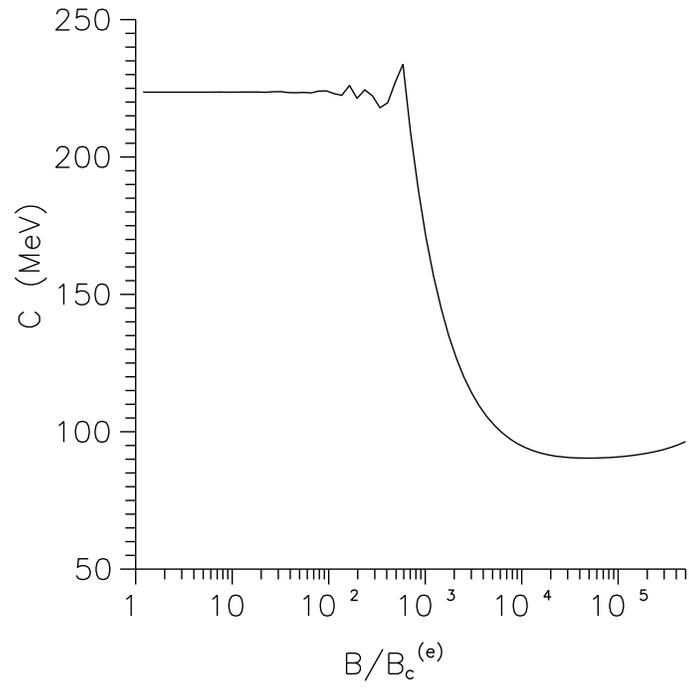,height=0.5\linewidth}
\caption{The variation of normalization constant $C$ with the magnetic field strength }
\end{figure}
\begin{figure}[ht]
\psfig{figure=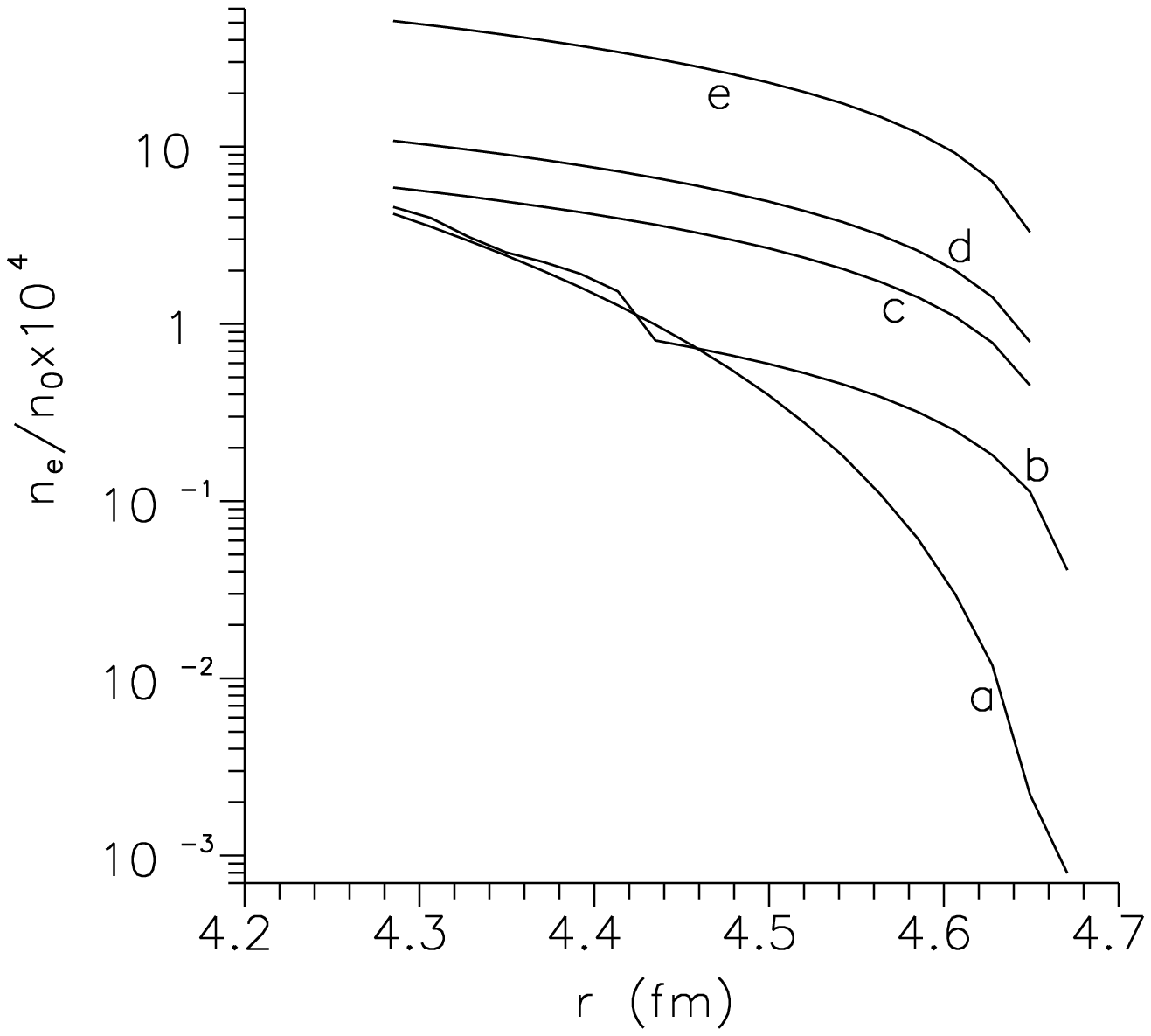,height=0.5\linewidth}
\caption{The variation of electron number density expressed in terms of normal nuclear density with
radial distance $r$ in Fermi. The curves indicated by $a$, $b$, $c$, $d$ and $e$ are for $B=10,
5\times 10^2, 5\times 10^3, 1\times 10^4$ and $5\times 10^4$ times $B_c^{(e)}$ respectively.}
\end{figure}
\begin{figure}[ht]
\psfig{figure=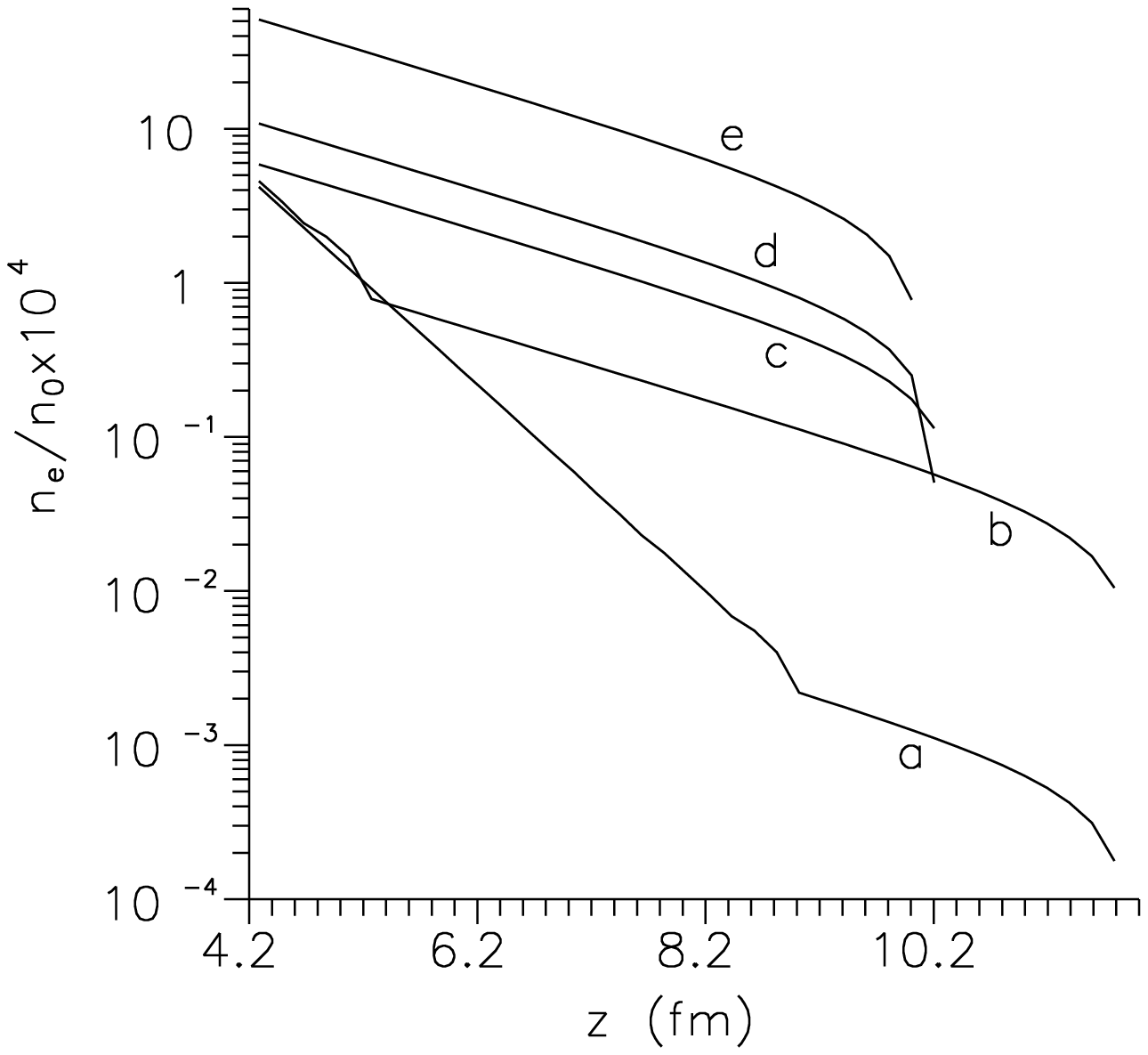,height=0.5\linewidth}
\caption{The variation of electron number density expressed in terms of normal nuclear density with
radial distance $z$ in Fermi. The curves indicated by $a$, $b$, $c$, $d$ and $e$ are for $B=10,
5\times 10^2, 5\times 10^3, 1\times 10^4$ and $5\times 10^4$ times $B_c^{(e)}$ respectively.}
\end{figure}
\begin{figure}[ht]
\psfig{figure=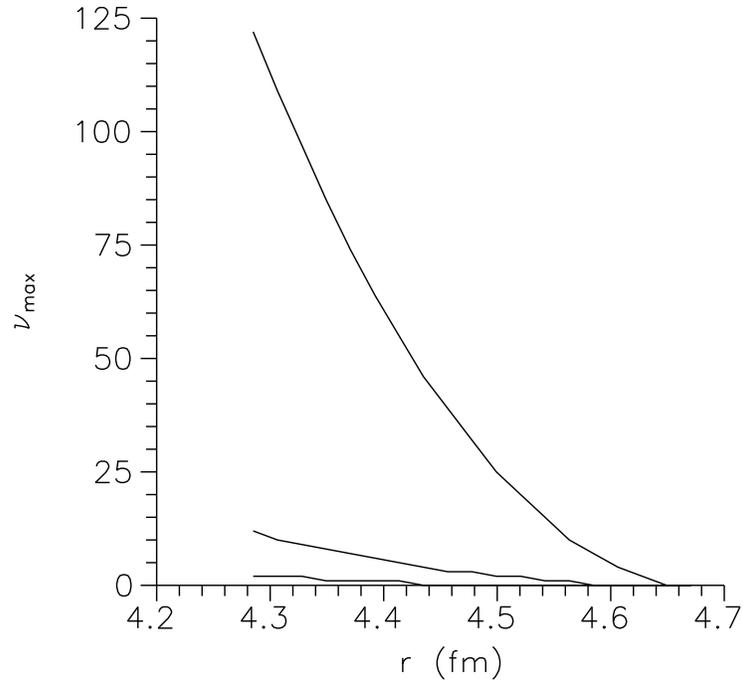,height=0.5\linewidth}
\caption{The variation of upper limit of Landau quantum number with the radial distance in Fermi.
Upper curve is for $B=10\times B_c^{(e)}$, middle one is for $10^2\times B_c^{(e)}$ and the lower one
is for $5\times 10^2 B_c^{(e)}$.}
\end{figure}
\begin{figure}[ht]
\psfig{figure=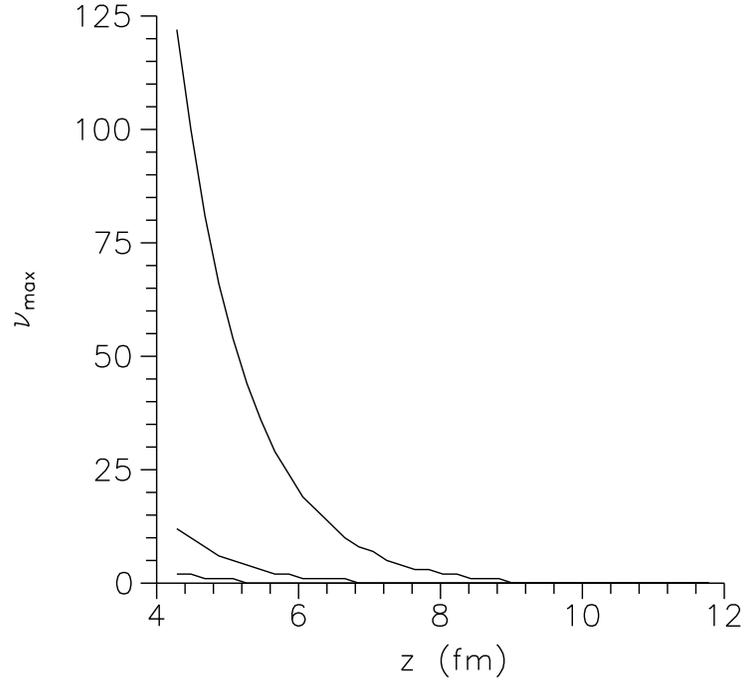,height=0.5\linewidth}
\caption{The variation of upper limit of Landau quantum number with the axial coordinate. Upper curve
is for $B=10\times B_c^{(e)}$, middle one is for $10^2\times B_c^{(e)}$ and the lower one
is for $5\times 10^2 B_c^{(e)}$.}
\end{figure}
\begin{figure}[ht]
\psfig{figure=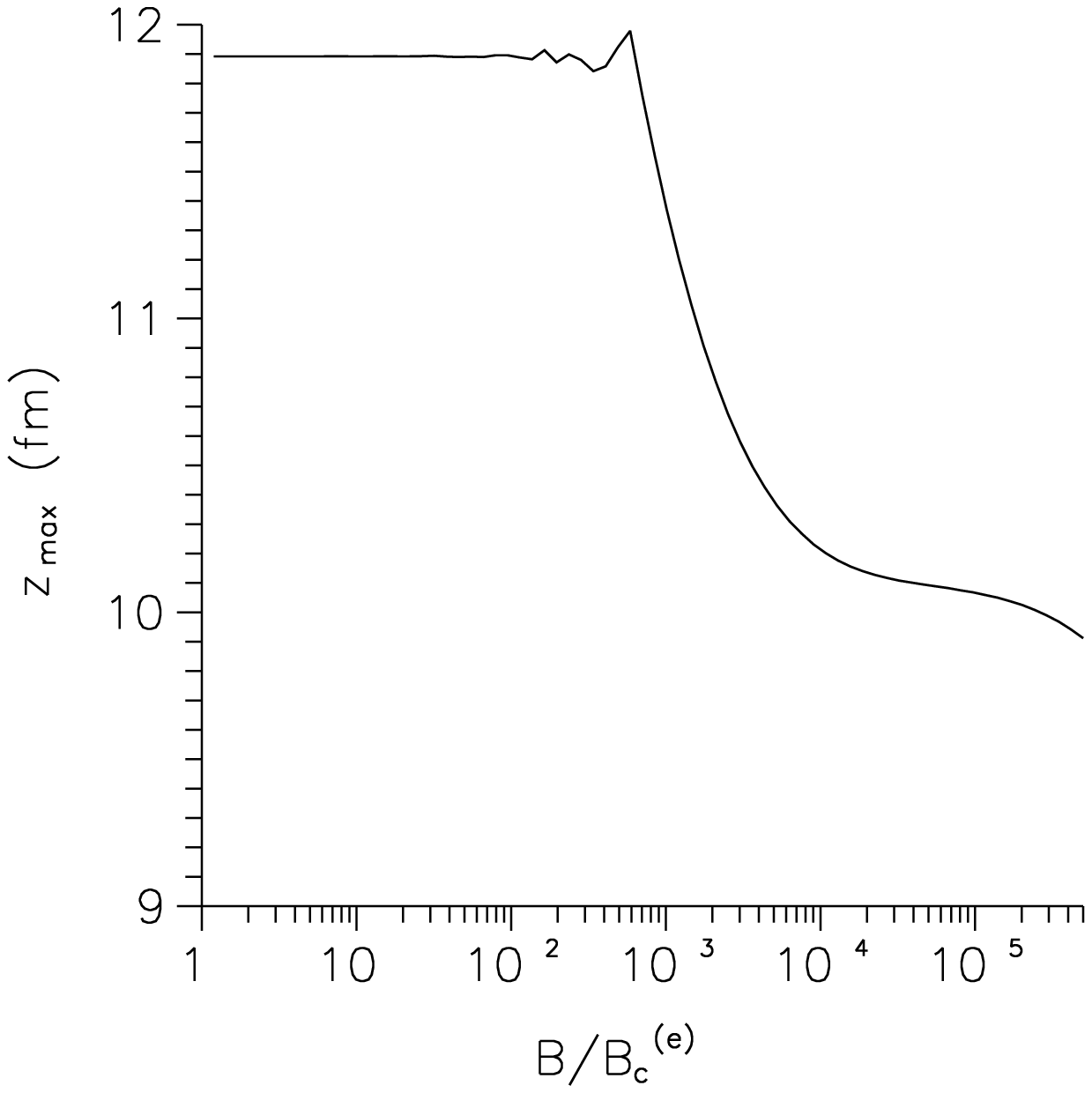,height=0.5\linewidth}
\caption{The variation of $z_{max}$ with the magnetic field strength.}
\end{figure}
\begin{figure}[ht]
\psfig{figure=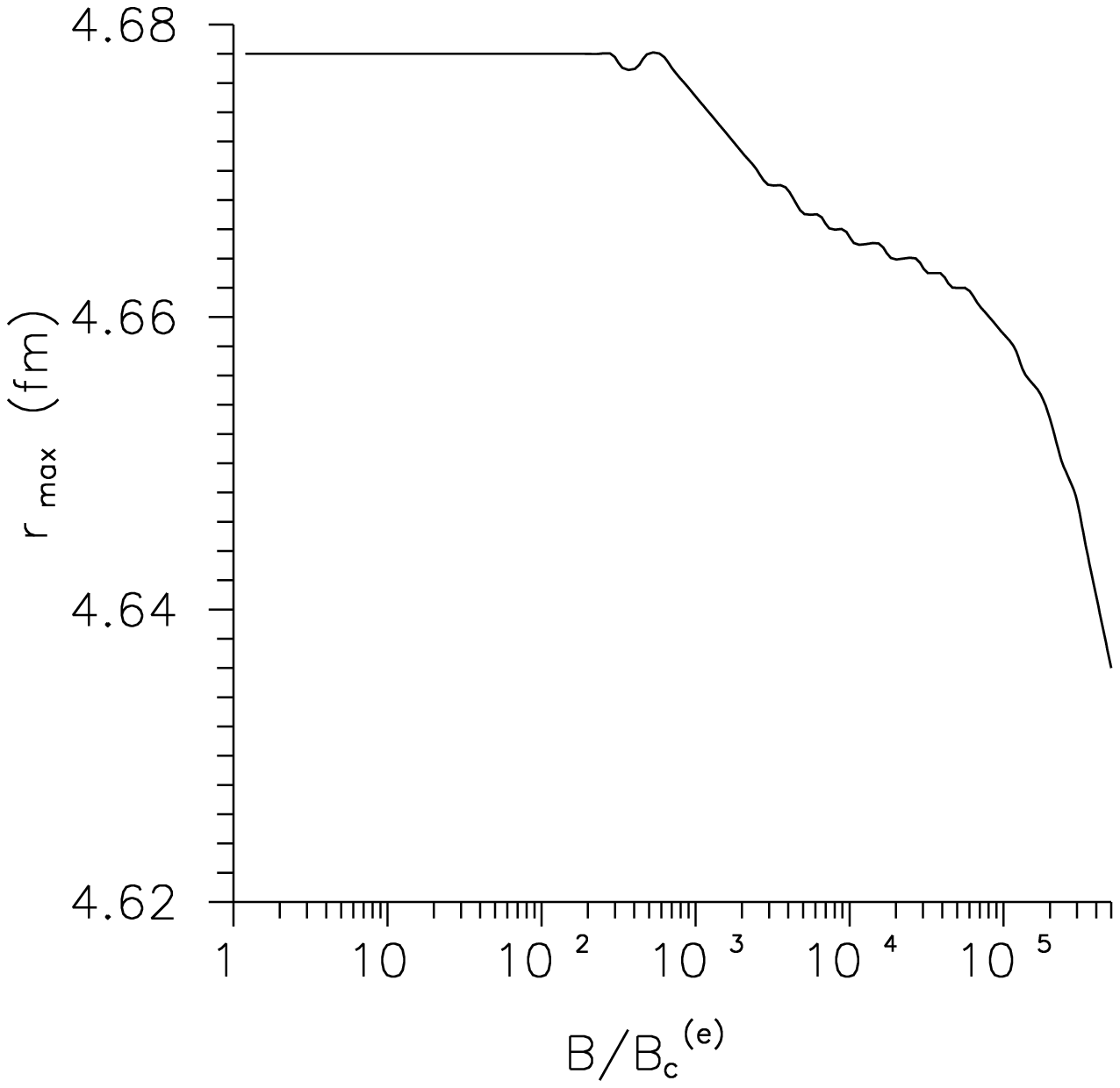,height=0.5\linewidth}
\caption{The variation of $r_{max}$ with the magnetic field strength.}
\end{figure}
\begin{figure}[ht]
\psfig{figure=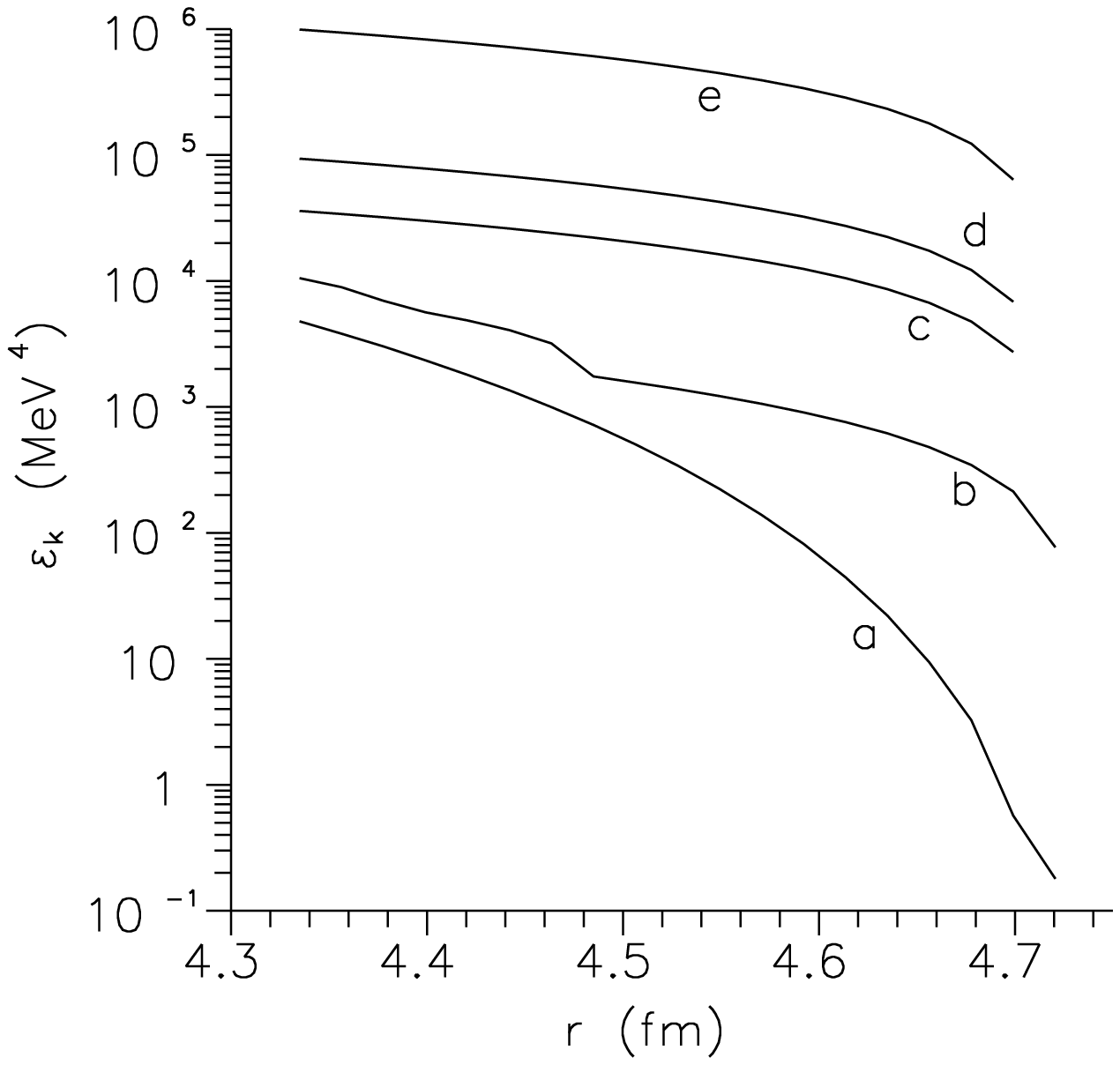,height=0.5\linewidth}
\caption{The variation of kinetic energy density with the radial distance in Fermi. The curves
indicated by $a$, $b$, $c$, $d$ and $e$ are for $B=10,
5\times 10^2, 5\times 10^3, 1\times 10^4$ and $5\times 10^4$ times $B_c^{(e)}$ respectively.}
\end{figure}
\begin{figure}[ht]
\psfig{figure=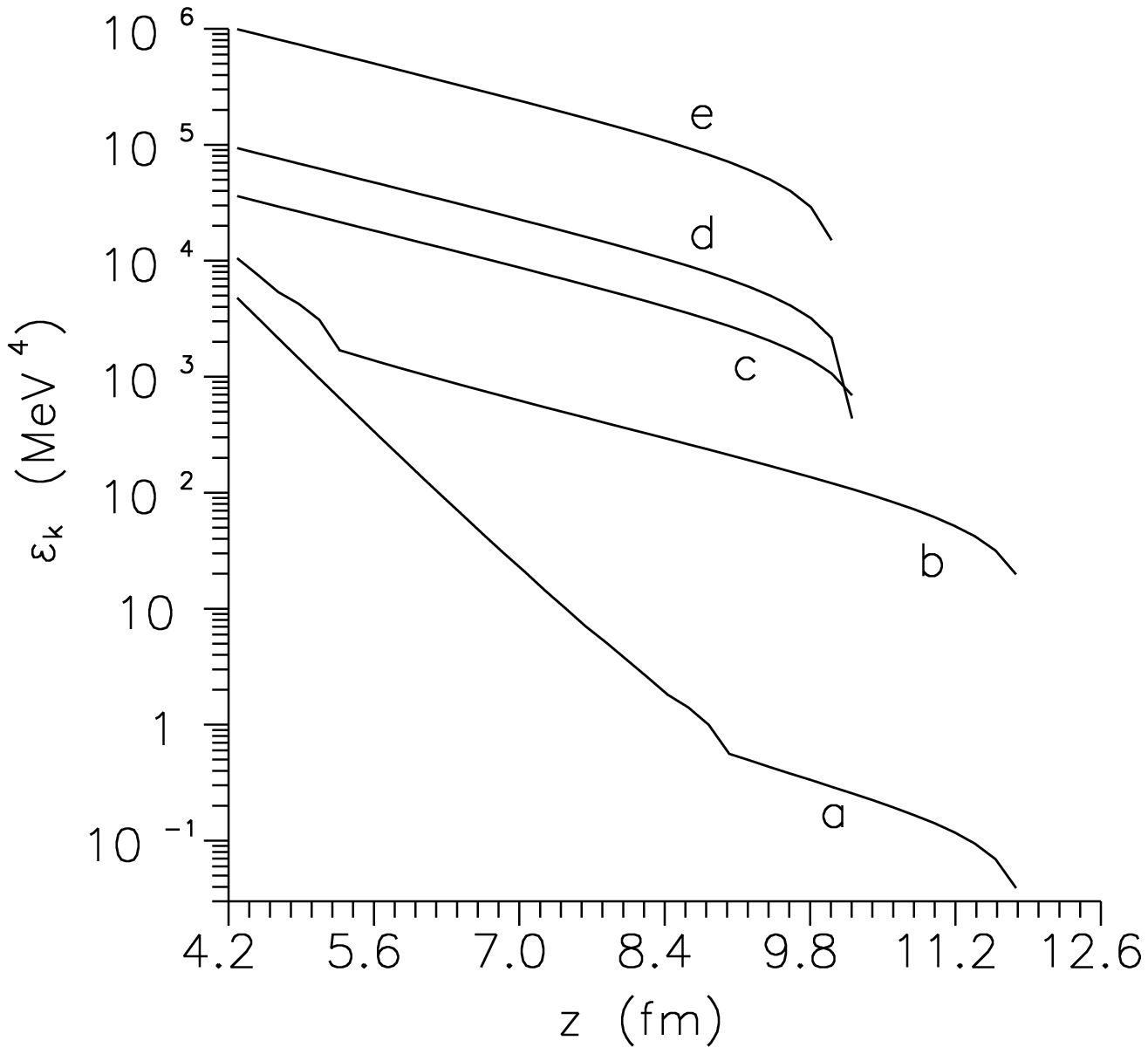,height=0.5\linewidth}
\caption{The variation of kinetic energy density with the axial coordinate in Fermi. The curves
indicated by $a$, $b$, $c$, $d$ and $e$ are for $B=10,
5\times 10^2, 5\times 10^3, 1\times 10^4$ and $5\times 10^4$ times $B_c^{(e)}$ respectively.}
\end{figure}
\begin{figure}[ht]
\psfig{figure=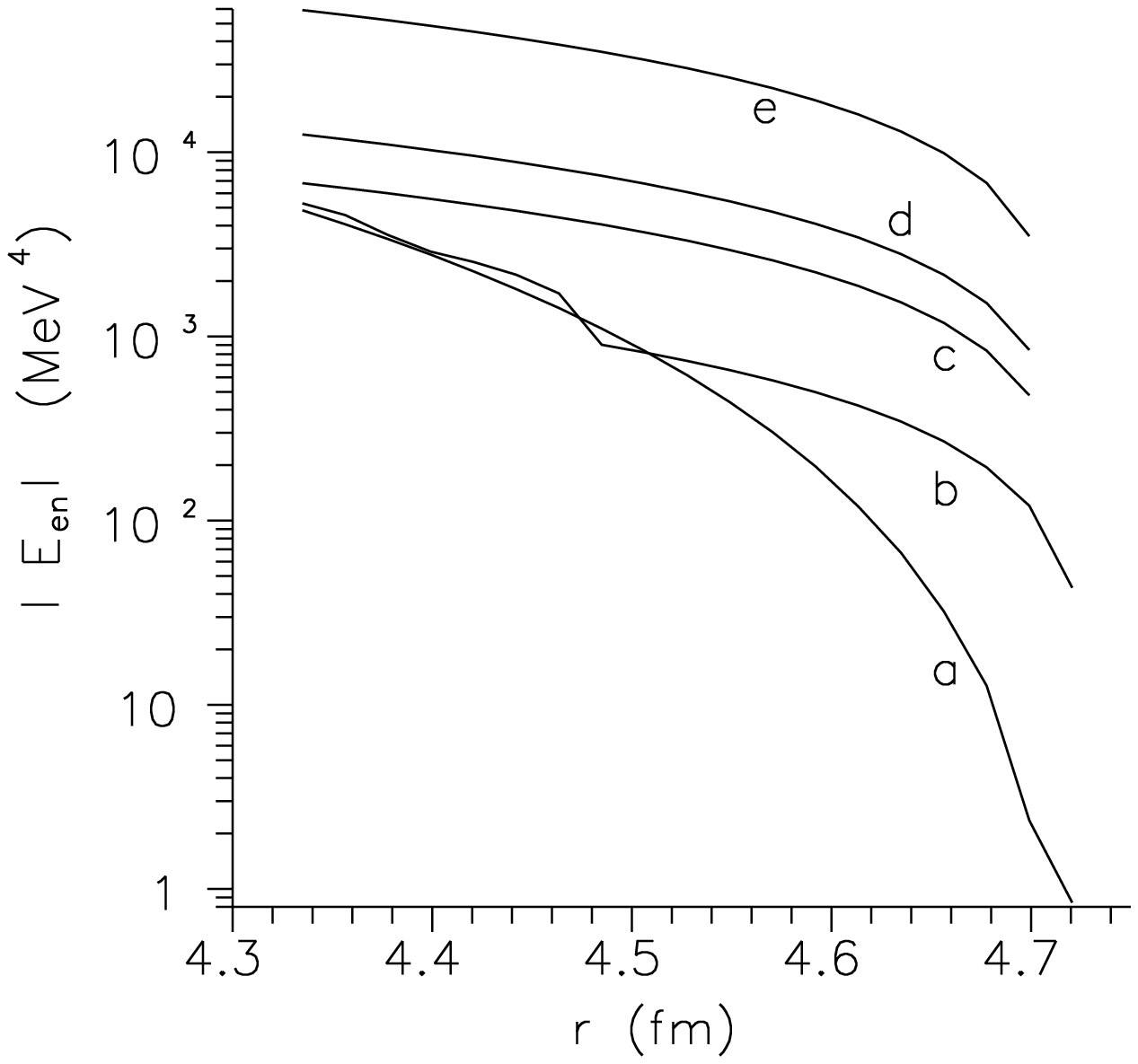,height=0.5\linewidth}
\caption{The variation of electron nucleus interaction energy density with the radial distance in
Fermi. The curves indicated by $a$, $b$, $c$, $d$ and $e$ are for $B=10,
5\times 10^2, 5\times 10^3, 1\times 10^4$ and $5\times 10^4$ times $B_c^{(e)}$ respectively.}
\end{figure}
\begin{figure}[ht]
\psfig{figure=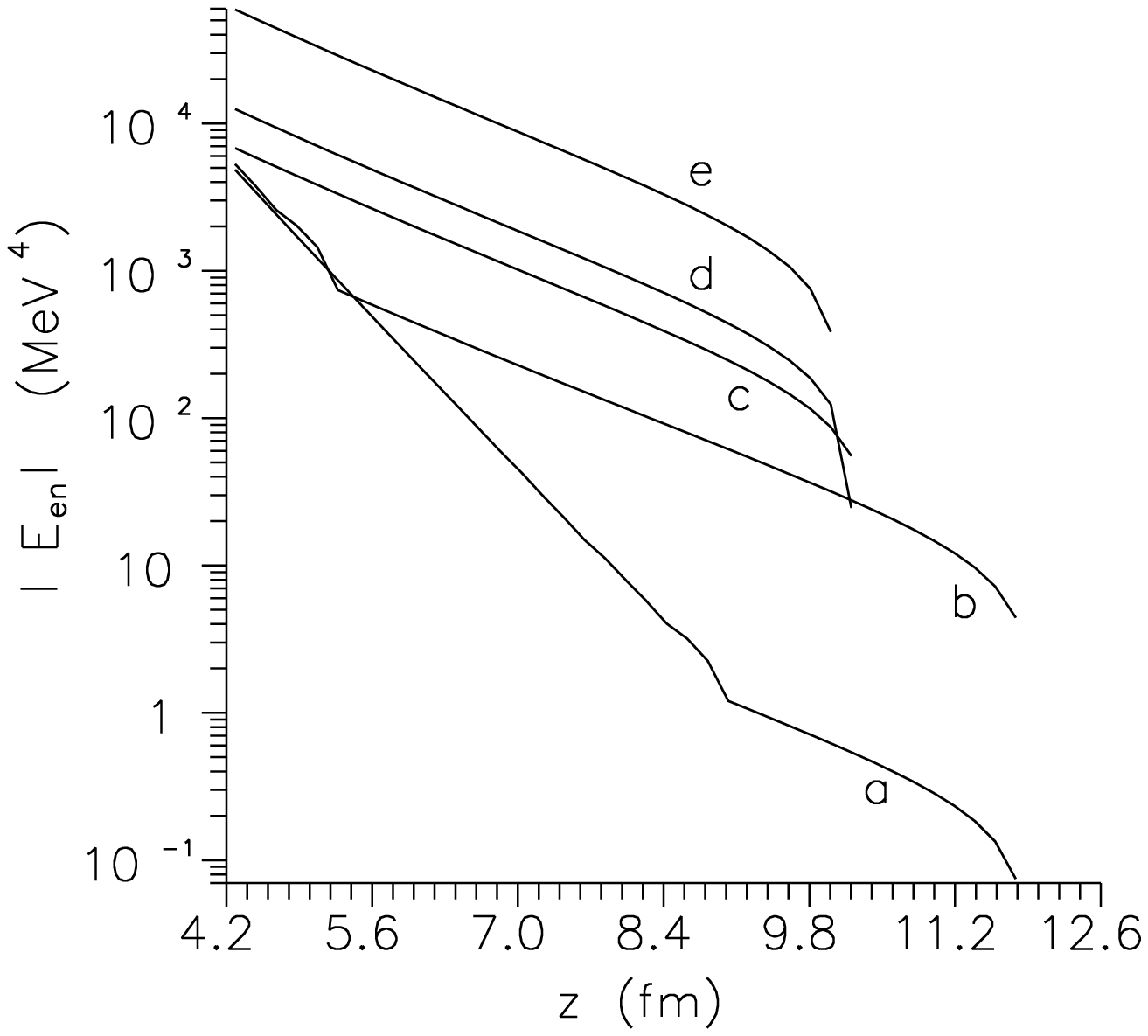,height=0.5\linewidth}
\caption{The variation of electron nucleus interaction energy density with the axial coordinate in
Fermi. The curves indicated by $a$, $b$, $c$, $d$ and $e$ are for $B=10,
5\times 10^2, 5\times 10^3, 1\times 10^4$ and $5\times 10^4$ times $B_c^{(e)}$ respectively.}
\end{figure}
\begin{figure}[ht]
\psfig{figure=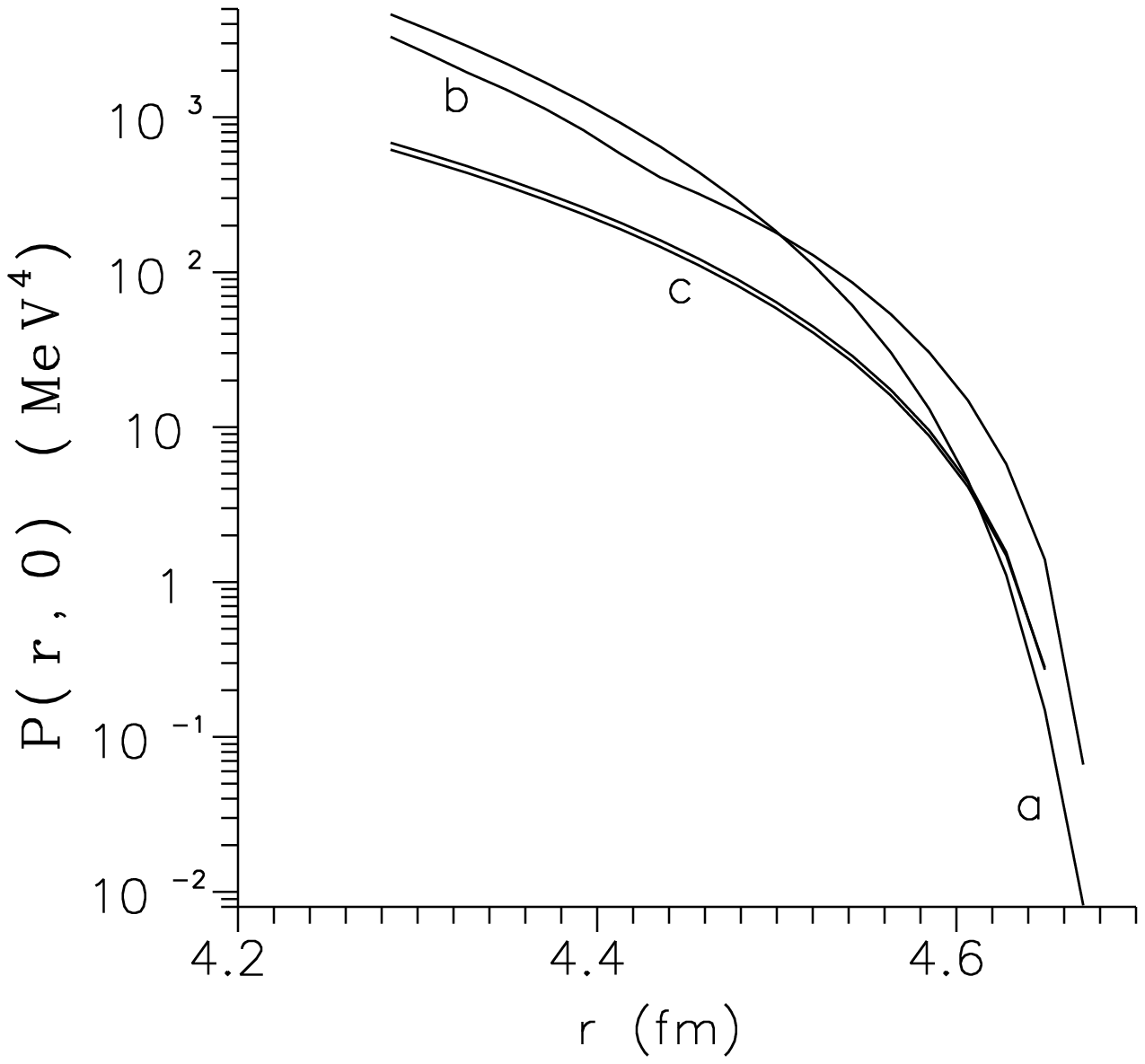,height=0.5\linewidth}
\caption{The variation of kinetic pressure of non-uniform electron gas within WS cell
with the radial distance in Fermi. Curves $a$ and $b$ are for $B=10\times B_c^{(e)}$ and $5\times
10^2B_c^{(e)}$ respectively, whereas upper and the lower curves indicated by $c$ are for $5\times
10^3B_c^{(e)}$ and $10^4\times B_c^{(e)}$ respectively.}
\end{figure}
\begin{figure}[ht]
\psfig{figure=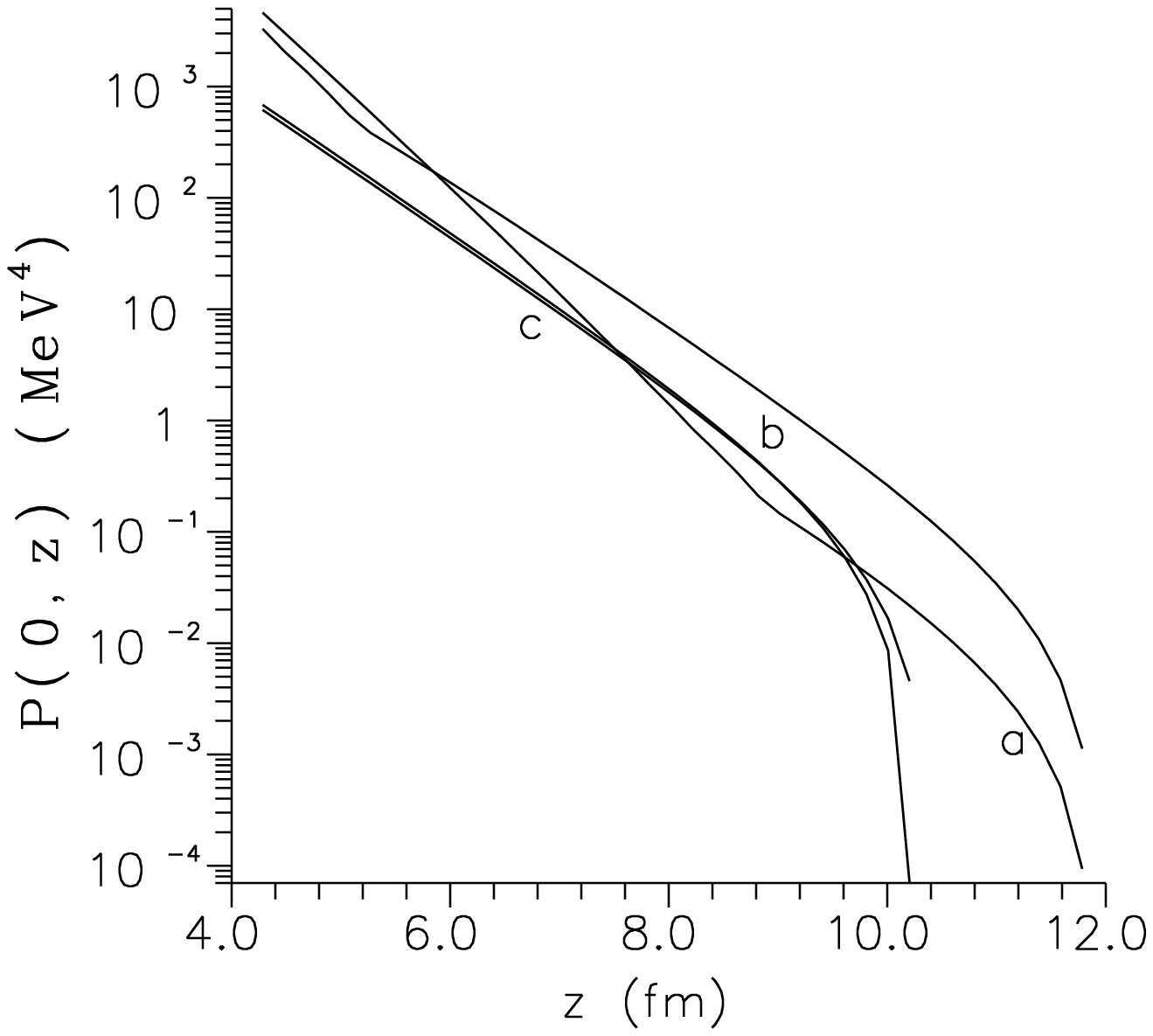,height=0.5\linewidth}
\caption{The variation of kinetic pressure of non-uniform electron gas within WS cell with the axial
coordinate in Fermi. Curves $a$ and $b$ are for $B=10\times B_c^{(e)}$ and $5\times
10^2B_c^{(e)}$ respectively, whereas upper and the lower curves indicated by $c$ are for $5\times
10^3B_c^{(e)}$ and $10^4\times B_c^{(e)}$ respectively.}
\end{figure}
\end{document}